\documentclass[journal]{IEEEtran} 
   \usepackage{graphicx}
\usepackage[linesnumbered,ruled]{algorithm2e}
\usepackage[noend]{algpseudocode}
\usepackage{booktabs}
\usepackage{amssymb}
\usepackage{amsthm}
\usepackage{multicol}
\usepackage{multirow} 
\usepackage{dsfont}
\usepackage[cmex10]{amsmath}
\usepackage{subcaption}
\usepackage{xcolor}
\usepackage[noadjust]{cite}

\begin{document}
\title{On Exact Distribution of Poisson-Voronoi Area in $K$-tier  HetNets with Generalized Association Rule}

\author{Washim Uddin Mondal and Goutam Das \thanks{The authors are with the G. S. Sanyal School of Telecommunications, IIT Kharagpur, India 721302.}}
        
\maketitle

\begin{abstract}
	This letter characterizes the exact distribution function of a typical  Voronoi area in a $K$-tier Poisson network. The users obey a generalized association (GA) rule, which is a superset of nearest base station association and maximum received power based association (with arbitrary fading) rules that are commonly adopted in the literature. Combining the Robbins' theorem and the probability generating functional of a Poisson point process, we obtain the \textit{exact} moments of a typical $k$-th tier Voronoi area, $k\in\{1,\dots,K\}$ under the GA rule. We apply this result in several special cases. For example, we prove that in multi-tier networks with the GA rule, the mean of $k$th tier Voronoi area can exactly be expressed in a closed-form. We also obtain simplified expressions of its higher-order moments for both average and instantaneous received power based user association. In single-tier networks with exponential fading, the later association rule provides closed-form expression of the second-order moment of a typical Voronoi area. We numerically evaluate this exact expression and compare it with an approximated result.
\end{abstract}
\begin{IEEEkeywords}
	Voronoi area distribution, Multi-tier PPP, Exact moments, Generalized association, Robbins' theorem.
\end{IEEEkeywords}

\IEEEpeerreviewmaketitle

\section{Introduction} 
\label{section_Introduction}


The Poisson point process (PPP) has popularly been utilized in the recent past as a mathematical model for a wide range of physical phenomena, ranging from species' habitat distribution in ecology \cite{renner2013equivalence} to the crystal growth in material science \cite{rios2009transformation}. In the field of wireless communications, this has been extensively used \cite{elsawy2013stochastic} to model the locations of base stations (BSs) and users in a given geographical area. One of the most prominent results that stems from this model is the simple analytical expressions of the ergodic capacity and the outage probability of downlink single-tier cellular networks \cite{andrews2011tractable}. This result can also be applied to the case of multi-tier heterogeneous networks (HetNets) \cite{dhillon2012modeling,heath2013modeling}, the device-to-device (D2D) networks \cite{ali2016modeling}, the vehicular ad-hoc networks \cite{farooq2015stochastic} and the Internet-of-Things (IoT) \cite{liu2019energy} network.

A key presumption in the above articles is that the potential transmitters are assumed to be \textit{always} active. Such supposition however, might not hold in a dense cellular network where the density of the deployed BSs is comparable to that of the users. Due to random placement of the users, in such a network, there will be a non-negligible probability, termed as \textit{void probability}, that an arbitrarily selected BS will not be associated with any user \cite{mondal2017uplink}. The void probability, therefore indicates the fraction of BSs that remains inactive. A plethora of articles have shown that the void probability significantly modifies the capacity and the outage probability of a cellular network \cite{liu2017limits}  and has major impact on its energy efficiency \cite{liu2016fundamentals}. Due to its  importance, an accurate characterization of the void probability has turned out to be an important goal in the current literature.

Note that that a $d$-dimensional PPP segregates the $\mathbb{R}^d$ space into infinitely many bounded Voronoi cells. Therefore, the void probability is essentially the probability that there are no users in the Voronoi cell associated with a typical BS. In particular,  if the  distribution function of a typical Voronoi area is indicated by $f(.)$ and $\lambda_0$ is the user density, then one can easily establish that the void probability, $\mathbb{P}_v(\lambda_0)$, can be obtained as:
\begin{align}
\label{Equation_Laplace}
\mathbb{P}_v(\lambda_0)\overset{(a)}{=}\int_{0}^{\infty} e^{-\lambda_0 A}f(A)\mathrm{d}A \overset{(b)}{=} \sum_{p=0}^{\infty} \dfrac{(-\lambda_0)^p}{p !}\mathbb{E}_{f(.)}[A^p]
\end{align}

where $\mathbb{E}_{f(.)}[A^p]$ is the $p$-th  order moment of the distribution function $f(.)$. Evidently the functions $\mathbb{P}_v(.)$ and $f(.)$ constitute a Laplace transform pair and thus have a $1-1$ correspondence. Characterization of $\mathbb{P}_v(.)$, hence is equivalent to the characterization of $f(.)$. Unfortunately, the \textit{exact} closed-form expression of $f(.)$, to the best of our knowledge, is still not available in the current literature. Nevertheless there has been  significant effort in the recent past to approximate $f(.)$ via extensive simulation. It has been reported that \cite{ferenc2007size}, for a two-dimensional PPP with density $\lambda$, the Voronoi cell area distribution, $f(.)$ can be well approximated by the following Gamma function:
\begin{align}
\label{Voronoi_area_approximation}
f(A)\approx \dfrac{(\zeta\lambda)^{\zeta}A^{\zeta-1}}{\Gamma(\zeta)}e^{-\zeta\lambda A}
\end{align}
where $\zeta = 7/2$. Substituting (\ref{Voronoi_area_approximation}) in relation $(a)$ in (\ref{Equation_Laplace}), one can obtain an approximate void probability as follows:
\begin{align}
\label{Approximate_Void_Prob}
\mathbb{P}_v(\lambda_0)\approx\left(1+\dfrac{\lambda_0}{\zeta \lambda}\right)^{-\zeta}
\end{align}

Eq. (\ref{Approximate_Void_Prob}) is used as a standard result in most of the literature. It is worth mentioning that the approximation (\ref{Approximate_Void_Prob}) is built on the presumption that the network under consideration is single-tier and the users adopt the nearest BS asociation rule. The relation (\ref{Approximate_Void_Prob}) gets slightly modified if either the network becomes multi-tiered \cite{liu2017limits} or the user association rule is changed \cite{liu2016fundamentals}. 

Parallel to these simulation-based-approximations, there has also been attempts to theoretically characterize the distribution function, $f(.)$. The main ingredient used in these line of attacks is the Robbins' theorem \cite{robbins1944measure}. It prescribes a numerical method to compute the moments of the Lebesgue measure of a random set. The moments  of $f(.)$ in a single-tier Poisson network (with closest BS association rule) can be \textit{exactly} obtained as a direct consequence of this theorem \cite{hayen2002areas}. Utilising the expansion $(b)$ in (\ref{Equation_Laplace}),
 the void probability, thus can be exactly evaluated.

In this article, we generalize this result to a $K$-tier Poisson network. Specifically, we assume each tier of BSs to be located according to an independent and homogeneous PPP. The users are assumed to follow a generalized association (GA) rule. The shortest distance based association and the maximum received power (with arbitrary fading distribution) based 
association are sub-cases of the GA rule. Within this framework, utilising the expression of probability
generating functional (PGFL) of  PPP and the  Robbins' theorem, we obtain  the exact expressions for arbitrary moments of a typical $k$-th tier Voronoi cell area where $ k \in\{1,\dots K\} $. Although the result is specifically derived for a two-dimensional PPP, it can be extended to higher dimensions as well. We prove that the main result gets simplified in several special cases. For example,

\textbullet ~A closed-form expression of the first order moment of the Voronoi area for a multi-tier network (with general association) is derived in the section \ref{section_first_order_moment}. The result is generalized for any arbitrary dimension. 

\textbullet~For the average received power based association rule, we derive a simplified expression of all the higher order moments of the Voronoi area in a $K$-tier environment. These expressions are further simplified for $K=1$ (section \ref{section_higher_order_moments}).

\textbullet~We also derive the simplified expressions of all the higher moments of the Voronoi area for instantaneous received power based association in an exponentially faded multi-tier network. For a single-tier network, this result provides us a closed-form expression of the second-order moment of its Voronoi area. We evaluate this expression numerically and compare the obtained value with an approximated result available in the literature.

\section{Preliminaries and System Model}
\label{section_prelim_system_model}
A $d$-dimensional homogeneous PPP, $\Phi$, with parameter $\lambda$ is characterized by the following two axioms.

\textbf{A}$_1$: For any set, $\mathcal{A}\subset\mathbb{R}^d$, one has:
\begin{align}
\label{Poisson_formula}
\mathbb{P}(\mathcal{N}(\Phi\cap\mathcal{A})=m)=e^{-\lambda A}\dfrac{(\lambda A)^m}{m!}, ~m=0,1,2,...
\end{align}

 where $\mathcal{N}(.)$ is the counting measure, and $A$ is the Lebesgue measure of $\mathcal{A}$. It is evident that the probability of $\Phi\cap\mathcal{A}$ being empty is $e^{-\lambda A}$. This justifies the expression of $\mathbb{P}_0(\lambda_0)$ in (\ref{Equation_Laplace}).

\textbf{A}$_2$: For any two disjoint subsets $\mathcal{A}$ and $\mathcal{B}$ of $\mathbb{R}^d$, the random variables $\mathcal{N}(\Phi \cap \mathcal{A})$ and $\mathcal{N}(\Phi \cap \mathcal{B})$ are independent.

In our paper we consider a $K$-tier downlink cellular network where the $k$-th tier BSs are deployed following a homogeneous $2$-dimensional PPP, $\Phi_k$, $k=1,..,K$ with density $\lambda_k$\footnote{The PPPs, $\{\Phi_k\}_{k=1}^{K}$ are assumed to be mutually independent.}. Another independent and homogeneous PPP $\Phi_0$ with density $\lambda_0$ depicts the location of the users. For any given realization of $\{\Phi_k\}_{k=1}^{K}$ let $\mathbf{r}_i$ denote the location of the $i$-th nearest BS from the origin. An arbitrarily selected user with location $\mathbf{x}_j$, will be associated with the $l$-th BS\footnote{The $l$-th nearest BS from the origin is simply referred to as the $l$-th BS.} if the following is true.
\begin{align}
\label{def_association_rule}
l=\underset{i\in\mathbb{N}}{\arg\max }~ w_{ij}|\mathbf{r}_i-\mathbf{x}_j|^{-\alpha}
\end{align}
where $\{w_{ij}\}_{i\in\mathbb{N}}$ denotes a set of independent random variables and $\alpha$ defines the pathloss coefficient. The association rule (\ref{def_association_rule}) is described as the generalized  association (GA) rule\footnote{A variant of this rule has appeared in \cite{liu2016fundamentals}.}. Various commonly used association rule such as the maximum average received power  based association (MARPA) and the maximum
instantaneous received power based  association (MIRPA) rules are special cases of the GA rule. For example, if the variables $w_{ij}$ are deterministic and defined as $w_{ij}=P_i$ where $P_i$ is the transmission power of the $i$-th BS, then the GA rule transforms into the MARPA rule\footnote{Observe that if the transmission powers of all the BSs are equal, then MARPA rule is equivalent into the nearest BS association rule.}. Also, if $w_{ij}=P_ih_{ij}$ where $\{h_{ij}\}_{i\in\mathbb{N}}$ is a collection of exponentially distributed, independent random variables with unit mean (denoting the small-scale fading), one obtains the MIRPA rule. We assume that the BSs are classified into multiple tiers based on the distributions of their associated weights $w_{ij}$. In particular,  $w_{ij}$'s corresponding to the $k$-th tier BSs, $k\in\{1,..,K\}$ are presumed to be distributed with density function $g_k(.)$ and cumulative function $G_k(.)$. The Voronoi cell $\mathcal{V}_l$,  associated with the $l$-th BS, is defined as follows:
\begin{align}
\mathcal{V}_l\triangleq \left\lbrace \mathbf{x}_j\in\mathbb{R}^2|~ l=\underset{i\in\mathbb{N}}{\arg\max }~ w_{ij}|\mathbf{r}_i-\mathbf{x}_j|^{-\alpha} \right\rbrace
\end{align}

\begin{figure*}[t!]
	\begin{equation}
	\label{Final_Epression}
	\begin{split}
	\mathbb{E}_{f_k(.)}[V_k^p] = \mathbb{E}_{\mathbf{w}_0^p}\left[\int_{(\mathbb{R}^2)^p} \exp\left(- \sum_{q=1}^{K}\int_{\mathbb{R}^2}\lambda_q\left[1-\prod_{j=1}^{p}G_q\left(w_{0j}\dfrac{|\mathbf{r}-\mathbf{x}_j|^{\alpha}}{|\mathbf{x}_j|^{\alpha}}\right)\right]\mathrm{d}\mathbf{r}\right)\mathrm{d}\mathbf{x}^p \right]
	\end{split}
	\end{equation}
	\hrulefill
\end{figure*}

Define $f_k(.)$ to be the  density function of a typical $k$-th tier Voronoi area. Our goal in this article is to exactly characterize the moments of $f_k$. We use two important results from random geometry to achieve this target. The first result is the Robbins' theorem \cite{robbins1944measure}. It states that, if $f_V(.)$ is the distribution function of $V$, the Lebesgue measure of a random subset $\mathcal{V}$ of $\mathbb{R}^d$, then  $\mathbb{E}_{f_V(.)}[V^p]$, the 
$p$-th order moment of $f_V(.)$ is given as follows:
\begin{align}
\label{Robbins_Theorem}
\mathbb{E}_{f_V(.)}[V^p] = \int_{\mathbb{R}^d}  \dots \int_{\mathbb{R}^d} \mathbb{P}\left(\mathbf{x}_1,\dots \mathbf{x}_p \in \mathcal{V}\right)\mathrm{d}\mathbf{x}_1\dots \mathrm{d}\mathbf{x}_p
\end{align}

The second result is the well-known PGFL expression of a PPP. Precisely, it states that, for a $d$-dimensional homogeneous PPP, $\Phi$, with density $\lambda$, the following holds true:
\begin{align}
\label{PGFL}
\mathbb{E}\left[\prod\limits_{\mathbf{x}_i\in\Phi\cap\mathcal{A}}s(\mathbf{x}_i)\right] = \exp\left(-\lambda\int_{\mathcal{A}}\left[1-s(\mathbf{x})\right]\mathrm{d}\mathbf{x} \right)
\end{align}
where $s(.)$ is any function and $\mathcal{A}\subset\mathbb{R}^d$. In the next section, we use these results to obtain arbitrary moments of the distribution functions, $f_k(.)$ of a typical $k$-th tier Voronoi area.

\section{The Main Result}
\label{section_main_result}

As homogeneous PPPs are stationary \cite{baccelli2009stochastic}, we, without loss of  generality, assume that a typical $k$-th tier BS,  $k\in\{1,\dots K\}$ is located at the origin. 
We represent its associated Voronoi cell as $\mathcal{V}_{\mathbf{0}}^k$. Eq. (\ref{Robbins_Theorem}) dictates that the $p$-th order moment of $f_k(.)$ can be derived by integrating $\mathbb{P}_{\mathbf{0}}(\mathbf{x}_1,\dots,\mathbf{x}_p\in\mathcal{V}_{\mathbf{0}}^k)$ over the region
 $(\mathbf{x}_1,\dots\mathbf{x}_p)\in(\mathbb{R}^2)^p$ 
 where
 $\mathbb{P}_{\mathbf{0}}(.)\triangleq \mathbb{P}(.|\mathbf{0}\in\Phi_k)$.
Let us select a point $\mathbf{x}_j$,  $j\in\{1,\dots, p\}$.
 The association rule (\ref{def_association_rule}) states that $\mathbf{x}_j\in \mathcal{V}_{\mathbf{0}}^k$ if the event, $\mathcal{E}_{ij}$ (described below) is true $\forall i\in\mathbb{N}$.
\begin{align}
\mathcal{E}_{ij}: w_{0j}|\mathbf{0}-\mathbf{x}_j|^{-\alpha}\geq w_{ij}|\mathbf{r}_i-\mathbf{x}_j|^{-\alpha}
\end{align}
where $\mathbf{r}_i\neq \mathbf{0}$ is the location of $i$-th closest BS to the origin, $\mathbf{0}$ and $\{w_{ij}\}_{i\in 0\cup \mathbb{N}}$ denotes independent random numbers. Define,  $\Phi_q^!  \triangleq \Phi_q/\{\mathbf{0}\}$, $q\in\{1,..., K\}$. One gets:
\begin{align}
\label{eq_11}
\mathbb{P}(\mathcal{E}_{ij}|\mathbf{r}_i\in\Phi_q^!,w_{0j}) = G_q\left(w_{0j}\dfrac{|\mathbf{r}_i-\mathbf{x}_j|^{\alpha}}{|\mathbf{x}_j|^{\alpha}}\right) 
\end{align}
where $G_q(.)$, as clarified in section \ref{section_prelim_system_model}, denotes the cumulative distribution function of the random weights associated with the $q$-th tier BSs.  Let $\mathbf{w}_0^p\triangleq \{w_{01},\dots w_{0p}\}$ and $\mathbf{x}^p \triangleq \{\mathbf{x}_1,\dots \mathbf{x}_p\}$. We have the following.
\begin{align}
\label{intermediate_result}
\begin{split}
&\mathbb{P}_{\mathbf{0}}(\mathbf{x}^p\in\mathcal{V}_{\mathbf{0}}^k|\mathbf{w}_0^p)=\mathbb{E}_{\mathbf{0}}\left[\prod_{q=1}^{K}\prod_{\mathbf{r}_i\in\Phi_q^!}\prod_{j=1}^{p} \mathbb{P}(\mathcal{E}_{ij}|\mathbf{r}_i\in\Phi_q^!,w_{0j})\right]\\
&\hspace{-0.1cm}\overset{(a)}{=}\prod_{q=1}^{K}\exp\left(-\lambda_q \int\limits_{\mathbb{R}^2}\left[1-\prod_{j=1}^{p}G_q\left(w_{0j}\dfrac{|\mathbf{r}-\mathbf{x}_j|^{\alpha}}{|\mathbf{x}_j|^{\alpha}}\right)\right]\mathrm{d}\mathbf{r}\right)
\end{split}
\end{align}
where $\mathbb{E}_{\mathbf{0}}[.]\triangleq \mathbb{E}[.|\mathbf{0}\in\Phi_k]$. Recall that  $\Phi_k,\Phi_q$ are  independent for $q\neq k$. Hence, $\mathbb{E}_{\mathbf{0}}[\prod_{q} h(\Phi_q^!)]$ $=\mathbb{E}_{\mathbf{0}}[h(\Phi_k^!)]\prod_{q\neq k}\mathbb{E}[h(\Phi_q^!)]$ for any function $h(.)$.
Moreover, due to the Slivnyak's theorem \cite{baccelli2009stochastic}, $\mathbb{E}_{\mathbf{0}}[h(\Phi_k^!)]=\mathbb{E}[h(\Phi_k)]$. 
Using this result, in conjunction with (\ref{PGFL}), (\ref{eq_11}), we derive  relation (a) in (\ref{intermediate_result}). 
Finally, we obtain: $\mathbb{P}_{\mathbf{0}}(\mathbf{x}^p\in \mathcal{V}_{\mathbf{0}}^k)=\mathbb{E}_{\mathbf{w}_0^p}\left[\mathbb{P}_\mathbf{0}{}(\mathbf{x}^p\in\mathcal{V}_{\mathbf{0}}^k|\mathbf{w}_0^p)\right]$.Note that the weights $\mathbf{w}_0^p$ are associated with the BS at the origin (which is presumed to belong to the $k$th tier) and hence are distributed with density function, $g_k(.)$. We now apply Robbins' Theorem (\ref{Robbins_Theorem}) to obtain $\mathbb{E}_{f_k(.)}[V_k^p]$, the $p$th order moment of $f_k(.)$. The final expression is given in (\ref{Final_Epression}). Although the result is derived for  $2$-dimensional  PPPs, it can be  extended to higher dimensions by  changing the space of integration of $\mathbf{r}$, $\mathbf{x}^p$ in (\ref{Final_Epression}). We examine the implication of (\ref{Final_Epression}) for some special cases in the section \ref{section_special_cases}.

\section{Special Cases}
\label{section_special_cases}

\subsection{First Order Moment}
\label{section_first_order_moment}

We first consider the first-order moment of $f_k(.)$. Note that, for $p=1$, the innermost integral of (\ref{Final_Epression}) is evaluated as follows:
\begin{align}
\label{special_case_1_1}
\begin{split}
&\int_{\mathbb{R}^2}\left[1-G_q\left(w_{01}\dfrac{|\mathbf{r}-\mathbf{x}_1|^{\alpha}}{|\mathbf{x}_1|^{\alpha}}\right)\right]\mathrm{d}\mathbf{r}\\
&\overset{(a)}{=}\int_{\mathbb{R}^2}\left[1-G_q\left(w_{01}\dfrac{|\mathbf{r}|^{\alpha}}{|\mathbf{x}_1|^{\alpha}}\right)\right]\mathrm{d}\mathbf{r}\\
&\overset{(b)}{=}\int_{0}^{\infty}\left[1-G_q\left(w_{01}\dfrac{{r}^{\alpha}}{|\mathbf{x}_1|^{\alpha}}\right)\right]{r}\mathrm{d}r\int_{0}^{2\pi}\mathrm{d}\theta\\
&\overset{(c)}{=}\dfrac{2\pi}{\alpha} |\mathbf{x}_1|^2 w_{01}^{-\frac{2}{\alpha}}\int_{0}^{\infty}\left[1-G_q(y)\right]y^{\frac{2}{\alpha}-1}\mathrm{d}y
\end{split}
\end{align}

where the relation $(a)$ is obtained by substituting $\mathbf{r}$ by $\mathbf{r}+\mathbf{x}_1$, $(b)$ by expanding the differential $\mathrm{d}\mathbf{r}$ as $r\mathrm{d}r\mathrm{d}\theta$ while $(c)$ follows from  substitution of $w_{01}r^{\alpha}/|\mathbf{x}_1|^{\alpha}$ by $y$. Define $\mathbb{E}_{q}[.]$ to be the expectation  over, $g_q(.)$, the density function of random weights associated with $q$-th tier BSs. We have:
\begin{align*}
\begin{split}
&\frac{2}{\alpha}\int\limits_{0}^{\infty}[1-G_q(y)]y^{\frac{2}{\alpha}-1}\mathrm{d}y
=\frac{2}{\alpha}\int\limits_{0}^{\infty}y^{\frac{2}{\alpha}-1}\int\limits_{y}^{\infty}g_q(w)\mathrm{d}w\mathrm{d}y
\\=&\frac{2}{\alpha}\int\limits_{0}^{\infty}g_q(w)\int\limits_{0}^{w}y^{\frac{2}{\alpha}-1}\mathrm{d}y\mathrm{d}w=\int\limits_{0}^{\infty}w^{\frac{2}{\alpha}}g_q(w)\mathrm{d}w=\mathbb{E}_q\left[W^{\frac{2}{\alpha}}\right]
\end{split}
\end{align*}
  
The final expression of (\ref{special_case_1_1}), hence, is: $\pi|\mathrm{x}_1|^2 w_{01}^{-\frac{2}{\alpha}}\mathbb{E}_q[W^{\frac{2}{\alpha}}]$. Substituting this result into (\ref{Final_Epression}), we obtain:
\begin{align}
\label{first_order_expression}
\begin{split}
&\mathbb{E}_{f_k(.)}[V_k]=\mathbb{E}_{k}\left[\int_{\mathbb{R}^2}e^{-\pi |\mathbf{x}_1|^2 w_{01}^{-\frac{2}{\alpha}}\sum_{q=1}^{K}\lambda_q\mathbb{E}_q[W^{\frac{2}{\alpha}}]}\mathrm{d}\mathbf{x}_1\right]\\
&\overset{(a)}{=} \mathbb{E}_{k}\left[\int_{0}^{\infty}e^{-\pi r^2 w_{01}^{-\frac{2}{\alpha}}\sum_{q=1}^{K}\lambda_q\mathbb{E}_q[W^{\frac{2}{\alpha}}]}r\mathrm{d}r\int_{0}^{2\pi}\mathrm{d}\theta\right]\\
&=\mathbb{E}_k[W^{\frac{2}{\alpha}}]\left(\sum_{q=1}^{K}\lambda_q\mathbb{E}_q[W^{\frac{2}{\alpha}}]\right)^{-1}
\end{split}
\end{align}

where $(a)$ is obtained by changing the integration into polar form. Equation (\ref{first_order_expression}) provides an exact closed-form expression of the first-order moment of a typical $k$th tier Voronoi cell area. The following points are worth mentioning. Firstly, for a single tier Poisson network with density $\lambda$, the mean area of a typical Voronoi cell is $1/\lambda$. Secondly, under both MARPA and MIRPA rule with exponentially faded channels (as described in section \ref{section_prelim_system_model}), the $k$-th tier mean Voronoi area is $P_k^{\frac{2}{\alpha}}/\sum_{q=1}^{K}\lambda_q P_q^{\frac{2}{\alpha}}$ where $P_q$ denotes the transmission power of $q$-th tier BSs. Finally the result (\ref{first_order_expression}) can also be generalized to higher dimension. For  $d$-dimensional PPPs we easily can demonstrate  that $\mathbb{E}_{f_k(.)}[V_k^p]=\mathbb{E}_k[W^{\frac{d}{\alpha}}]/\sum_{q=1}^{K}\lambda_q\mathbb{E}_q[W^{\frac{d}{\alpha}}]$. We will now discuss  higher order  moments of $f_k(.)$.

\subsection{Higher Order Moments}
\label{section_higher_order_moments}

Unlike the mean, it is in general difficult to obtain simplified closed-form expressions of the higher order moments of $f_k(.)$. Hence, one must resort to numerical methods to evaluate these quantities. However in some special cases partial simplification can be obtained. Below we discuss two such specific scenarios.

We first consider a multi-tier network with the MARPA rule i.e. $w_{ij}=P_q$ if $\mathbf{r}_i\in\Phi_q$ and $P_q$ defines the transmission power of $q$-th tier BSs.
In this case,  $G_q(w)=\mathds{1}(w\geq P_q)$ where $\mathds{1}(.)$ is the indicator function. Observe that,
\begin{align}
\label{expression_Gq}
\begin{split}
&\int_{\mathbb{R}^2}\left[1-\prod_{j=1}^{p}G_q\left(P_k\dfrac{|\mathbf{r}-\mathbf{x}_j|^{\alpha}}{|\mathbf{x}_j|^{\alpha}}\right)\right]\mathrm{d}\mathbf{r}\\
&=\int_{\mathbb{R}^2}\left[1-\prod_{j=1}^{p}\mathds{1}\left(|\mathbf{r}-\mathbf{x}_j|^2\geq \left(\dfrac{P_q}{P_k}\right)^{\frac{2}{\alpha}}|\mathbf{x}_j|^2\right)\right]\mathrm{d}\mathbf{r}\\
&=\int_{\mathbb{R}^2}\left[1-\mathds{1}\left(\mathbf{r}\not\in\bigcap\limits_{j=1}^{p}\mathcal{C}\left(\mathbf{x}_j,(P_q/P_k)^{\frac{1}{\alpha}}|\mathbf{x}_j|\right)\right)\right]\mathrm{d}\mathbf{r}\\
&=\int_{\mathbb{R}^2}\mathds{1}\left(\mathbf{r}\in\bigcup\limits_{j=1}^{p}\mathcal{C}\left(\mathbf{x}_j,(P_q/P_k)^{\frac{1}{\alpha}}|\mathbf{x}_j|\right)\right)\mathrm{d}\mathbf{r}
\end{split}
\end{align}
where $\mathcal{C}(\mathbf{x},r)$ denotes the set of interior points of a circle with centre at $\mathbf{x}$ and radius $r$. It is obvious that $|\mathcal{A}|=\int_{} \mathds{1}(\mathbf{r}\in\mathcal{A})\mathrm{d}\mathbf{r}$ for any deterministic set $\mathcal{A}\subset \mathbb{R}^d$ where the integration is over  $\mathbb{R}^d$ and $|.|$ is the Lebesgue measure\footnote{With slight abuse of the notations, we use $|.|$ to indicate both the Lebesgue measure of a set and the norm of a vector.}. Thus, (\ref{Final_Epression}) reduces to:
\begin{align}
\label{pMoment_MARPA}
\begin{split}
\mathbb{E}_{f_k(.)}[V_k^p]=\int\limits_{(\mathbb{R}^2)^p} e^{-\sum\limits_{q=1}^{K}\lambda_q\big|\bigcup\limits_{j=1}^{p}\mathcal{C}\left(\mathbf{x}_j,(P_q/P_k)^{\frac{1}{\alpha}}|\mathbf{x}_j|\right)\big|}\mathrm{d}\mathbf{x}^p
\end{split}
\end{align}
 
 For single-tier PPPs with density $\lambda$,  (\ref{pMoment_MARPA}) reduces to:
 \begin{align}
 \label{pMoment_MARPA_single_tier}
 	\mathbb{E}_{f(.)}[V^p]=\int_{(\mathbb{R}^2)^p}\exp\left(-\lambda\bigg|\bigcup\limits_{j=1}^{p}\mathcal{C}(\mathbf{x}_j,|\mathbf{x}_j|)\bigg|\right)\mathrm{d}\mathbf{x}^p
 \end{align}
 
 It is worth pointing out that the expression (\ref{pMoment_MARPA_single_tier}) can be derived via a simpler method. Recall that, in single-tier networks,
 MARPA is equivalent to the nearest BS association rule. Thus, the event that the point $\mathbf{x}_1$ associates itself with the BS located at the origin (denoted as BS$_0$) is identical to the event that there are no BSs nearer to $\mathbf{x}_1$ than BS$_0$. This indicates that the disk,
 $\mathcal{C}(\mathbf{x}_1,|\mathbf{x}_1|)$ must be devoid of any BSs. 
 The  probability of this event is $e^{-\lambda|\mathcal{C}(\mathbf{x_1},|\mathbf{x}_1|)|}$. In a similar manner, the probability that $p$ arbitrary points $(\mathbf{x}_1,\dots,\mathbf{x}_p)$ associate themselves with BS$_0$ is $\exp(-\lambda\bigcup \mathcal{C}(\mathbf{x}_j,|\mathbf{x}_j|))$. Applying Robbin's formula, one can now obtain (\ref{pMoment_MARPA_single_tier}).
 Details of this process, along with numerical evaluation of (\ref{pMoment_MARPA_single_tier}), for $p=2,3$, can be found in \cite{hayen2002areas}.

 We now consider the MIRPA rule in an exponentially faded multi-tier environment. Particularly, $G_q(w)=1-e^{-\mu_qw}$, $w\geq 0$ for some parameter $\mu_q>0$. We have,
 \begin{align}
 \label{pMoment_Intermediate_result_MIRPA}
 \begin{split}
 &\int_{\mathbb{R}^2}\left[1-\prod_{j=1}^{p}G_q\left(w_{0j}\dfrac{|\mathbf{r}-\mathbf{x}_j|^{\alpha}}{|\mathbf{x}_j|^{\alpha}}\right)\right]\mathrm{d}\mathbf{r}=\sum_{s=1}^{p}(-1)^{s+1}\\
 &\times\sum_{\substack{\mathcal{J}\subset\{1,\dots,p\}\\ |\mathcal{J}|=s}}\underbrace{\int_{\mathbb{R}^2}\exp\left(-\mu_q\sum_{j\in \mathcal{J}}w_{0j}\dfrac{|\mathbf{r}-\mathbf{x}_j|^{\alpha}}{|\mathbf{x}_j|^{\alpha}}\right)\mathrm{d}\mathbf{r}}_{H_{\mathcal{J}}^q}
 \end{split}
 \end{align}
 
 The terms $H_{\mathcal{J}}^q$ given in (\ref{pMoment_Intermediate_result_MIRPA}), in general need to be computed numerically. However, for $\alpha=2$, its closed-form expressions can be derived. Define $\beta_j\triangleq w_{0j}|\mathbf{x}_j|^{-2}$. We get,
 \begin{align}
 \label{rearrangement_square_sum}
 \begin{split}
 &\sum_{j\in\mathcal{J}}w_{0j}\dfrac{|\mathbf{r}-\mathbf{x}_j|^2}{|\mathbf{x}_j|^2}=\sum_{j\in\mathcal{J}} \beta_j\Big\lbrace|\mathbf{r}|^2 -2(\mathbf{r}\cdot \mathbf{x}_j) + |\mathbf{x}_j|^2\Big\rbrace\\
 =&\left(\sum_{j\in\mathcal{J}}\beta_j\right)|\mathbf{r}|^2 -  2\left(\sum_{j\in\mathcal{J}} \beta_j \mathbf{x}_j\right)\cdot \mathbf{r} + \sum_{j\in\mathcal{J}} w_{0j}\\
 =&\left[\sum_{j\in\mathcal{J}}\beta_j\right]\Bigg|\mathbf{r}-\dfrac{ \sum\limits_{j\in\mathcal{J}}\beta_{j}\mathbf{x}_j}{\sum\limits_{j\in\mathcal{J}}\beta_j}\Bigg|^2-\dfrac{\Big| \sum\limits_{j\in\mathcal{J}}\beta_{j}\mathbf{x}_j\Big|^2}{\sum\limits_{j\in\mathcal{J}}\beta_j}+\sum\limits_{j\in\mathcal{J}}w_{0j}
 \end{split}
 \end{align}
 
 Injecting (\ref{rearrangement_square_sum}) into (\ref{pMoment_Intermediate_result_MIRPA}), it is easy to show that,
 \begin{align}
 \label{def_H_q_J}
 H_{\mathcal{J}}^q=\dfrac{\pi\mu_q^{-1}}{\sum\limits_{j\in\mathcal{J}}\beta_j}\exp\left(-\mu_q\sum\limits_{j\in\mathcal{J}}w_{0j}+\mu_q\dfrac{\Big| \sum\limits_{j\in\mathcal{J}}\beta_{j}\mathbf{x}_j\Big|^2}{\sum\limits_{j\in\mathcal{J}}\beta_j}\right)
 \end{align}
 
 The expression of  $\mathbb{E}_{f_k(.)}[V_k^p]$, as suggested in equation (\ref{Final_Epression}), therefore can be equivalently written as the following.
 \begin{align}
 \label{pMoment_final}
 \begin{split}
 &\mathbb{E}_{f_k(.)}[V_k^p]=
 \int\limits_{(\mathbb{R}^+)^p}\mathrm{d}\mathbf{w}_0^p(\mu_k)^p\exp\left(-\mu_k\sum_{j=1}^{p}w_{0j}\right)\times\\
 &\int\limits_{(\mathbb{R}^2)^p}\mathrm{d}\mathbf{x}^p\exp\left(-\sum_{s=1}^p(-1)^{s+1}\sum_{\substack{\mathcal{J}\subset\{1,\dots,p\}\\ |\mathcal{J}|=s}}\sum_{q=1}^{K}\lambda_qH_{\mathcal{J}}^q\right) 
 \end{split}
  \end{align}
  
  This result can be further simplified for $p=2$ (second-order moment) in a single-tier network with density $\lambda$ and associated weight distribution $G(w)=1-e^{-\mu w}$, $w\geq 0$. In this scenario, 
  \begin{align}
  \label{2ndMoment_FinalExpression}
  \begin{split}
  \mathbb{E}_{f(.)}[V^2]= \dfrac{1}{\lambda^2}\Bigg[\sum_{k=0}^{\infty}\dfrac{B(k+1)}{(k+1)}+\dfrac{2k^2 B(k+2)}{(k+1)^2}\Bigg]
  \end{split}
  \end{align}
  where $B(m)\triangleq \int_{0}^{1}t^{m-1}(1-t)^{m-1}\mathrm{d} t$ defines the Beta function with identical arguments. Eq. (\ref{2ndMoment_FinalExpression}) is derived in the Appendix. 
  
  It would be interesting to see how different approximations, such as given in \cite{liu2016fundamentals}, compares with the exact moment derived in (\ref{2ndMoment_FinalExpression}). Numerical evaluation of (\ref{2ndMoment_FinalExpression}) dictates that, $\mathbb{E}_{f(.)}[V^2]=$ $1.122/\lambda^2$. On the other hand, \cite{liu2016fundamentals} suggests that for an arbitrary pathloss coefficient $\alpha$, the Voronoi area distribution in a single-tier network with  density $\lambda$ can accurately be approximated by  (\ref{Voronoi_area_approximation}) with the following  parameter: $\zeta= \frac{7}{2}\mathbb{E}[W^{\frac{2}{\alpha}}]\mathbb{E}[W^{-\frac{2}{\alpha}}]$. Note that for exponentially distributed weights $\zeta=\frac{7}{2}\Gamma(1+\frac{2}{\alpha})\Gamma(1-\frac{2}{\alpha})$ where $\Gamma(.)$ indicates the Gamma function. It is evident that $\zeta\rightarrow \infty$ as $\alpha\rightarrow 2$.
   In this limiting scenario, the approximation in \cite{liu2016fundamentals} provides  
    $\mathbb{E}_{f(.)}[V^2]={\lambda^{-2}}$ which is  approximately $12\%$ erroneous as compared to its exact value.
       \section{Conclusion}
   
   We derive the exact expressions of the moments of a typical Voronoi area in a multi-tier Poisson network with a generalized association rule. This fully characterizes the area distribution function of a Voronoi cell and the base station void probability. We use this result in several special cases and obtain simplified expressions of the moments. For example, we derive a closed-form expression of the first-order moment of the Voronoi area in $K$-tier network with generalized association and the second-order moment in a single-tier network with exponentially faded channels under  maximum instantaneous received power based user association rules. Numerical computation reveals that the approximations available in the literature may produce as large as $12\%$ error as compared to the exact results obtained via our method. This necessitates a better approximation result for the area distribution function of the Voronoi cells for a wide range of system parameters. We expect that our exact result will help in achieving such a goal in the future.

  \appendix
  \section*{Derivation of (\ref{2ndMoment_FinalExpression})}
  \label{appendix_derivation_2nd_Moment}

  It is evident from (\ref{pMoment_final}) that, 
  \begin{align}
  \label{2MomentExpression}
  \mathbb{E}_{f(.)}[V^2]=\mathbb{E}_{\mathbf{w}_0^2}\left[\int_{(\mathbb{R}^2)^2}e^{-\lambda(H_1+H_2-H_{1,2})}\mathrm{d}\mathbf{x}^2\right]
  \end{align}
  
   Using (\ref{def_H_q_J}), one gets $H_j=\pi/(\mu \beta_{j})$,  where $\beta_j\triangleq w_{0j}/|\mathbf{x}_j|^2$, $j\in\{1,2\}$. In a similar fashion, $H_{1,2}$ $=\pi
   [\mu(\beta_1+\beta_2)]^{-1}$ $e^{-H_0}$ where $H_0=[\mu(w_{01}\beta_2+w_{02}\beta_1)$ $-2\mu\beta_1\beta_2(\mathbf{x}_1\cdot \mathbf{x}_2)]/(\beta_1+\beta_2)$. The term $\mathbf{x}_1\cdot \mathbf{x}_2$ is the dot product between $\mathbf{x}_1$ and $\mathbf{x}_2$. Writing $|\mathbf{x}_j|$ as $x_j$, the differential $\mathrm{d}\mathbf{x}_j$ as $x_j\mathrm{d}x_j\mathrm{d}\theta_j$ for $j\in\{1,2\}$ and $\mathbf{x}_1\cdot \mathbf{x}_2 = x_1 x_2 \cos(\theta_1-\theta_2)$, one can write the integration given in (\ref{2MomentExpression}) over $(\mathbf{x}_1,\mathbf{x}_2)\in (\mathbb{R}^2)^2$ as an equivalent integration over $(x_1,x_2,\theta_1,\theta_2)\in [0,\infty)^2\times [0,2\pi]^2$. Moreover, substituting $x_1$ as $a_1r\cos\phi$, $x_2$ as $a_2r\sin\phi$ and $\mathrm{d}x_1 \mathrm{d}x_2$ as $a_1a_2r\mathrm{d}r\mathrm{d}\phi$ where $a_j=(\mu w_{0j}/\pi \lambda)^{\frac{1}{2}}$, $j\in \{1,2\}$, the integration over $(x_1,x_2)\in$ $  [0,\infty)^2$ can be changed into an integration over $(r,\theta)\in [0,\infty)$ $\times[0,\pi/2]$. The final result is the following.
\begin{align}
\begin{split}
\label{2ndMomentIntermediate}
&J\triangleq\int_{(\mathbb{R}^2)^2}e^{-\lambda(H_1+H_2-H_{1,2})}\mathrm{d}\mathbf{x}^2=\dfrac{\mu^2 w_{01}w_{02}}{\pi^2\lambda^2} \times
\\
&\int_{[0,2\pi]^2}\int_{0}^{\frac{\pi}{2}}\int_{0}^{\infty} F(r,\phi,\theta_1-\theta_2)r^3e^{-r^2}\mathrm{d}r\mathrm{d}\phi\mathrm{d}\theta_1\mathrm{d}\theta_2
\end{split}
\end{align}

where the function $F$ is expressed as follows:
\begin{align*}
\begin{split}
F&(r,\phi,\theta)\triangleq\sin\phi \cos\phi\\
\times& \exp\Big[r^2(\sin\phi\cos\phi)^2 e^{-F_1(\phi)}e^{-F_2(\phi)}e^{F_3(\phi,\theta)}\Big]\\
=&\sum_{k=0}^{\infty}\dfrac{r^{2k}}{k!}(\sin\phi \cos\phi)^{2k+1} e^{-kF_1(\phi)}e^{-kF_2(\phi)}e^{kF_3(\phi,\theta)}\\
=&\sum_{k,t=0}^{\infty}\dfrac{r^{2k}}{k!}(\sin\phi\cos\phi)^{2k+1}e^{-k[F_1(\phi)+F_2(\phi)]} \dfrac{k^tF_3^t(\phi,\theta)}{t!}
\end{split}
\end{align*}
where $F_1(\phi)=\mu w_{01}\cos^2\phi$, $F_2(\phi)=\mu w_{02}\sin^2\phi$ and finally, $F_3(\phi,\theta)=2\mu\sqrt{w_{01}w_{02}}\sin\phi\cos\phi\cos(\theta)$. Injecting the series expansion of $F(r,\phi,\theta)$ in (\ref{2ndMomentIntermediate}), the term $J$  can be written as:
\begin{align}
\label{IntermediateR}
\begin{split}
&J=\sum_{k=0}^{\infty}\sum_{t=0}^{\infty} \dfrac{k^t}{\lambda^2} (\mu^2 w_{01}w_{02})^{\frac{1}{2}t+1} \underbrace{\int_{0}^{\infty}\dfrac{r^{2k+3}}{k!}e^{-r^2}\mathrm{d}r}_{J_1}\\
&\times \underbrace{\dfrac{1}{\pi^2 t!}\int_{[0,2\pi]^2}2^t\cos^t(\theta_1-\theta_2)\mathrm{d}\theta_1\mathrm{d}\theta_2}_{J_2}\\
&\times \int_{0}^{\frac{\pi}{2}}e^{-kF_1(\phi)}e^{-kF_2(\phi)}(\sin\phi\cos\phi)^{2k+t+1}\mathrm{d}\phi
\end{split}
\end{align}  
  
Substituting $r$ by $\sqrt{y}$ in $J_1$, it can be shown that,
\begin{align}
J_1=\dfrac{1}{2k!}\int_{0}^{\infty}y^{k+1}e^{-y}\mathrm{d}y=\dfrac{1}{2k!}\Gamma(k+2)=\dfrac{1}{2}(k+1)
\end{align} 
where $\Gamma(.)$ is the Gamma function. Now we calculate $J_2$. Note that $J_2=0$ for odd $t$. Presume, $t=2q$ for some integer $q\geq 0$. In this case, we have the following.
\begin{align*}
\begin{split}
 J_2&=\dfrac{2\pi4^q}{\pi^2(2q)!}\int_{0}^{2\pi}\cos^{2q}\theta \mathrm{d}\theta=\dfrac{8\pi4^q}{\pi^2(2q)!}\int_{0}^{\frac{\pi}{2}}\cos^{2q}\theta\mathrm{d}\theta\\
&=\dfrac{4\pi4^{q}}{\pi^2(2q)!} \dfrac{\Gamma\left(q+\dfrac{1}{2}\right)\Gamma\left(\dfrac{1}{2}\right)}{\Gamma(q+1)}\\
&=\dfrac{4\pi4^{q}}{\pi^2(2q)!}\dfrac{\left(\dfrac{2q-1}{2}\right)\left(\dfrac{2q-3}{2}\right)\dots\dfrac{1}{2}\sqrt{\pi}\sqrt{\pi}}{q!}=\dfrac{4}{(q!)^2}
\end{split}
\end{align*}
    
Using these results, (\ref{IntermediateR}) can be re-written as:
\begin{align}
\label{Expression_J_final}
\begin{split}
J=&\dfrac{2}{\lambda^2}\sum_{k=0}^{\infty}\sum_{q=0}^{\infty} (k+1)k^{2q} \int_{0}^{\frac{\pi}{2}} (\sin\phi\cos\phi)^{2k+2q+1}\\
&\times \dfrac{(\mu w_{01})^{q+1}}{q!}e^{-kF_1(\phi)} \dfrac{(\mu w_{02})^{q+1}}{q!}e^{-kF_2(\phi)}\mathrm{d} \phi
\end{split}
\end{align}

Now, $w_{01},w_{02}$ are exponentially distributed with parameter $\mu$. Therefore, for any integer $q$, we have:
\begin{align}
\label{Moments_of_Exponential}
\begin{split}
\mathbb{E}[(\mu w_{0j})^{q+1}e^{-kF_j(\phi)}]= \dfrac{(q+1)!}{(1+kQ_j(\phi))^{q+2}}, j\in\{1,2\}
\end{split}
\end{align} 
where $Q_1(\phi)=\cos^2(\phi), Q_2(\phi)=\sin^2(\phi)$. We thus obtain:
 \begin{align}
 \label{Expectation_J}
 \begin{split}
 \mathbb{E}_{\mathbf{w}_0^2}[J]
 =&\dfrac{2}{\lambda^2}\sum_{k=0}^{\infty}\int_{0}^{\frac{\pi}{2}}\dfrac{(k+1)(\sin\phi\cos\phi)^{2k+1}}{(1+k\cos^2\phi)^2(1+k\sin^2\phi)^2}\\
 \times \sum_{q=0}^{\infty}&(q+1)^2 \left[\dfrac{k^2\sin^2\phi\cos^2\phi}{(1+k\cos^2\phi)(1+k\sin^2\phi)}\right]^q \mathrm{d}\phi
 \end{split}
 \end{align}
 
 It is easy to prove that, $\sum_{q=0}^{\infty}(q+1)^2x^q=(1+x)/(1-x)^3$, $|x|<1$. Using this result in (\ref{Expectation_J}) and simplifying the resulting expression, we can arrive at (\ref{2ndMoment_FinalExpression}).
  


\ifCLASSOPTIONcaptionsoff
  \newpage
\fi

\bibliographystyle{IEEEtran}
\bibliography{Bib}

\begin{thebibliography}{10}
\providecommand{\url}[1]{#1}
\csname url@samestyle\endcsname
\providecommand{\newblock}{\relax}
\providecommand{\bibinfo}[2]{#2}
\providecommand{\BIBentrySTDinterwordspacing}{\spaceskip=0pt\relax}
\providecommand{\BIBentryALTinterwordstretchfactor}{4}
\providecommand{\BIBentryALTinterwordspacing}{\spaceskip=\fontdimen2\font plus
\BIBentryALTinterwordstretchfactor\fontdimen3\font minus
  \fontdimen4\font\relax}
\providecommand{\BIBforeignlanguage}[2]{{%
\expandafter\ifx\csname l@#1\endcsname\relax
\typeout{** WARNING: IEEEtran.bst: No hyphenation pattern has been}%
\typeout{** loaded for the language `#1'. Using the pattern for}%
\typeout{** the default language instead.}%
\else
\language=\csname l@#1\endcsname
\fi
#2}}
\providecommand{\BIBdecl}{\relax}
\BIBdecl

\bibitem{renner2013equivalence}
I.~W. Renner and D.~I. Warton, ``Equivalence of {MAXENT} and {P}oisson point
  process models for species distribution modeling in ecology,''
  \emph{Biometrics}, vol.~69, no.~1, pp. 274--281, 2013.

\bibitem{rios2009transformation}
P.~Rios and E.~Villa, ``Transformation kinetics for inhomogeneous nucleation,''
  \emph{Acta Materialia}, vol.~57, no.~4, pp. 1199--1208, 2009.

\bibitem{elsawy2013stochastic}
H.~ElSawy, E.~Hossain, and M.~Haenggi, ``Stochastic geometry for modeling,
  analysis, and design of multi-tier and cognitive cellular wireless networks:
  A survey,'' \emph{IEEE Communications Surveys \& Tutorials}, vol.~15, no.~3,
  pp. 996--1019, 2013.

\bibitem{andrews2011tractable}
J.~G. Andrews, F.~Baccelli, and R.~K. Ganti, ``A tractable approach to coverage
  and rate in cellular networks,'' \emph{IEEE Trans Commun}, vol.~59, no.~11,
  pp. 3122--3134, 2011.

\bibitem{dhillon2012modeling}
H.~S. Dhillon, R.~K. Ganti, F.~Baccelli, and J.~G. Andrews, ``Modeling and
  analysis of k-tier downlink heterogeneous cellular networks,'' \emph{IEEE J.
  Sel. Areas Commun.}, vol.~30, no.~3, pp. 550--560, 2012.

\bibitem{heath2013modeling}
R.~W. Heath, M.~Kountouris, and T.~Bai, ``Modeling heterogeneous network
  interference using {P}oisson point processes,'' \emph{IEEE Trans. Signal
  Process.}, vol.~61, no.~16, pp. 4114--4126, 2013.

\bibitem{ali2016modeling}
K.~S. Ali, H.~ElSawy, and M.-S. Alouini, ``Modeling cellular networks with
  full-duplex {D}2{D} communication: A stochastic geometry approach,''
  \emph{IEEE Trans. Commun.}, vol.~64, no.~10, pp. 4409--4424, 2016.

\bibitem{farooq2015stochastic}
M.~J. Farooq, H.~ElSawy, and M.-S. Alouini, ``A stochastic geometry model for
  multi-hop highway vehicular communication,'' \emph{IEEE Trans. Wireless
  Commun.}, vol.~15, no.~3, pp. 2276--2291, 2015.

\bibitem{liu2019energy}
C.-H. Liu, Y.-H. Shen, and C.-H. Lee, ``Energy-efficient activation and uplink
  transmission for cellular {I}o{T},'' \emph{IEEE Internet Things J.}, 2019.

\bibitem{mondal2017uplink}
W.~U. Mondal and G.~Das, ``Uplink user process in {P}oisson cellular network,''
  \emph{IEEE Commun. Lett.}, vol.~21, no.~9, pp. 2013--2016, 2017.

\bibitem{liu2017limits}
C.-H. Liu and H.-C. Tsai, ``On the limits of coexisting coverage and capacity
  in multi-rat heterogeneous networks,'' \emph{IEEE Trans. Wireless Commun.},
  vol.~16, no.~5, pp. 3086--3101, 2017.

\bibitem{liu2016fundamentals}
C.-H. Liu and K.~L. Fong, ``Fundamentals of the downlink green coverage and
  energy efficiency in heterogeneous networks,'' \emph{IEEE J. Sel. Areas
  Commun.}, vol.~34, no.~12, pp. 3271--3287, 2016.

\bibitem{ferenc2007size}
J.-S. Ferenc and Z.~N{\'e}da, ``On the size distribution of {P}oisson {V}oronoi
  cells,'' \emph{Physica A: Statistical Mechanics and its Applications}, vol.
  385, no.~2, pp. 518--526, 2007.

\bibitem{robbins1944measure}
H.~E. Robbins, ``On the measure of a random set,'' \emph{The Annals of
  Mathematical Statistics}, vol.~15, no.~1, pp. 70--74, 1944.

\bibitem{hayen2002areas}
A.~Hayen and M.~Quine, ``Areas of components of a {V}oronoi polygon in a
  homogeneous {P}oisson process in the plane,'' \emph{Advances in Applied
  Probability}, vol.~34, no.~2, pp. 281--291, 2002.

\bibitem{baccelli2009stochastic}
F.~Baccelli and B.~Blaszczyszyn, ``Stochastic geometry and wireless networks,
  part {I}: Theory,'' in \emph{Now Publishers}, 2009.

\end{thebibliography}

\end{document}